\documentclass[conference,10pt]{IEEEtran}
\usepackage{epsfig,makeidx,color}
\usepackage{amsmath,amssymb,bbm}
\usepackage{cite,graphicx,lipsum}
\usepackage{enumerate}
\usepackage[switch,pagewise]{lineno}
\usepackage{hyperref}
\hypersetup{
        colorlinks = true,
        citecolor=blue,
}
\def\uR{{\mathbb R}}

\def\be{ \begin{equation} }
\def\ee{ \end{equation} }
\def\bea{ \begin{eqnarray} }
\def\eea{ \end{eqnarray} }

\def\b0{{\bf 0}}

\ifCLASSOPTIONonecolumn
  \newcommand{\figwidth}{0.50\columnwidth}
  
\else
  \newcommand{\figwidth}{0.85\columnwidth}
  
\fi

\IEEEoverridecommandlockouts

\title{IoT-based  Analysis for Smart Energy Management}

\author{\IEEEauthorblockN{Guang-Li Huang, Adnan Anwar, Seng W. Loke, Arkady Zaslavsky, 
and Jinho Choi} 
\IEEEauthorblockA{School of Information Technology, Deakin University, Geelong, Australia\\
\{belinda.huang,adnan.anwar,seng.loke,arkady.zaslavsky,jinho.choi\}@deakin.edu.au}
\thanks{The Centre for New Energy Technologies, “C4NET” has contributed to the funding of this project. C4NET acknowledges the major funding contribution of the
Victorian Department of Environment, Land, Water and
Planning and its Core Participants. The views expressed herein are not necessarily the views
of C4NET or the Victorian State Government and neither
party accepts responsibility for any information or advice
contained herein.}}

\begin{document}

\maketitle

\begin{abstract}
Smart energy management based on the Internet of Things (IoT) aims to achieve optimal energy utilization through real-time energy monitoring and analyses of power consumption patterns in IoT networks (e.g., residential homes and offices) supported by wireless technologies - this is of great significance for the sustainable development of energy. Energy disaggregation is an important technology to realize smart energy management, as it can determine the power consumption of each appliance from the total load (e.g., aggregated data). Also, it gives us clear insights into users' daily power-consumption-related behaviours, which can enhance their awareness of power-saving and lead them to a more sustainable lifestyle. This paper reviews the state-of-the-art algorithms for energy/power disaggregation and public datasets of power consumption. Also,  potential use cases for smart energy management based on IoT networks are presented 
along with a discussion of open issues for future study.

\end{abstract}
\ifCLASSOPTIONonecolumn
\baselineskip 26pt
\fi

\section{Introduction}\label{sec:1}

The sustainable use of energy is one of the biggest challenges that our society is facing today \cite{kolter2011redd}. As the demand for energy (e.g., electricity) continues growing, it would be significant if we could use and manage energy in a smarter way (i.e., smart energy management). Statistically, the main power use can be divided into two major categories: commercial and residential electricity consumption, namely, mainly lighting and using household devices. 
Since a considerable portion of the power consumption is closely related to people’s indoor activities, such as cooking with a microwave or oven in the kitchen, using a computer/laptop or printer at work, watching TV or listening to music as entertainment, heating, cooling, and so on, monitoring the energy consumption, analyzing power usage patterns and recognizing the residents electricity-consumption behaviors will help better energy management.

To effectively analyze power usage patterns for optimal energy utilization, the premise is that the power consumption of each device can be accurately obtained. Early statistics-based methods collected the data by deploying measurement equipment to each device to enable fine-grained energy monitoring. Due to the requirement for pre-installation and extra cost, this kind of method is not popular. The Non-intrusive Load Monitoring (NILM) technology began to attract wide attention as early as the 1990s \cite{henriet2018generative} \cite{hart1992nonintrusive},  which aims to determine the energy consumption of each individual appliance based on the detailed analysis of the total load. That is, the characteristics of the appliance (e.g., the real and reactive power) is used as unique ``signatures" for each appliance, and devices can be discern from the aggregated data, namely, energy disaggregation \cite{zoha2012non}. 

The Internet-of-Things (IoT) is to connect devices and sensors through the Internet and there are a number of IoT applications such as smart cities, smart factory, and so on \cite{Asghari19}. To support wireless connectivity for things (i.e., devices and sensors), various technologies can be  considered with short-range and long-range connectivity \cite{Ding_20Access}. In most IoT applications,  while sensors and devices are expected to transmit their measurements (e.g., thermometer sensors send their temperature readings), they can also be used to send power consumption data (e.g., smart plugs). Thus, IoT networks become capable of supporting smart energy management through data analytics \cite{Marjani17}. In particular, the
energy disaggregation used in individual homes or offices can be performed simultaneously on a large-scale in tens or hundreds of homes or offices via IoT networks, which raises the possibility of   large-scale smart energy management, e.g., over neighbourhoods, or suburbs.

In this paper, we review the state-of-the-art algorithms for energy/power disaggregation and public datasets of power consumption, present potential use case scenarios for smart energy management based on IoT networks, and discuss challenges and open issues for future research.

The rest of the paper is organized as follows. Section \ref{sec:2} introduces the high-level framework of IoT networks for smart energy management. Section \ref{sec:3} presents energy disaggregation in terms of concepts, algorithms, and analysis of energy usage patterns. Section \ref{sec:4} summarizes the public datasets used for energy /power disaggregation. Section \ref{sec:5} proposes potential use-case scenarios and discusses open issues. Section \ref{sec:6} concludes the paper.

\section{Smart Energy Management}\label{sec:2}

\begin{figure}[ht]
\begin{center}
\includegraphics[width=\figwidth]{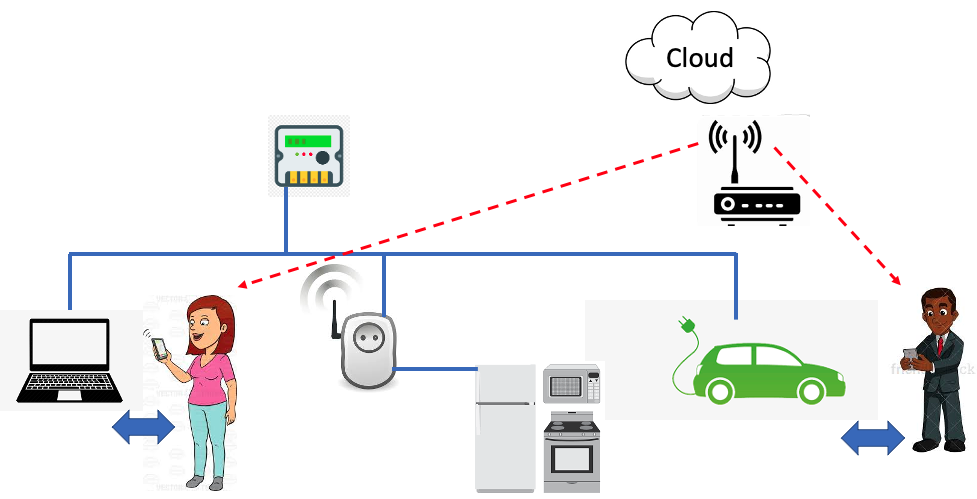} 
\end{center}
\caption{An illustration of IoT network for smart energy management.}
        \label{Fig:system}
\end{figure}

IoT networks can help collect datasets from appliances and devices at homes and offices to understand power consumption patterns by individuals and guide users for smart energy usage. As shown in Fig.~\ref{Fig:system}, we assume that an IoT network is deployed at a home and a number of appliances and devices are connected to a local home server (directly or indirectly through smart plugs), which is able to perform data analytic and train models to understand power consumption patterns.

We assume that different technologies can be used for wireless connectivity. For example, WiFi, Zigbee, or bluetooth can be used. The roles of the local home server for smart energy management are as follows:
\begin{enumerate}[i.]
    \item Collecting real-time power consumption data from appliances and devices 
    \item Training machine learning (ML) models to perform power disaggregation and user detection 
    \item Recommending smart energy usage for users
\end{enumerate}
Note that for the third role, the local home server needs to access the cloud to find the energy cost at a given time and duration so that it can provide a recommendation to individuals to lower the energy cost.
If all the appliances and devices are directly connected with  with wireless transceivers to the local home server, their real-time power consumption information can be known by the local home server. However, there might be some appliances without  wireless transceivers. Thus, we can consider the use of smart plugs (or a smart meter) that can provide the power consumption information for any connected appliances and devices. Since a smart plug (or a smart meter) may collect the aggregated power consumption, the local home server needs to perform energy disaggregation,
which will be explained in the following section.

\section{Energy Disaggregation}\label{sec:3}

\subsection{An Overview of Energy Disaggregation} 

Energy disaggregation is a challenging blind source separation problem that infer energy consumption of electrical appliances on each branch circuit in a building through total power consumption \cite{jazizadeh2018embed}, aiming to separate the energy consumption of individual appliances from the total consumption readings of multiple appliances (i.e., the aggregated data at the smart meter) \cite{Jiang21}. Fig.~1 is presented to illustrate the overall concept of energy disaggregation of individual appliances. In the figure, the end-user individual appliances data are sent to a local aggregator which is later sent to the edge or cloud for energy management including operation and retail billing. In a traditional smart home concept, the smart meters are responsible to capture the aggregated information and communicate with the operators. However, in an edge/IoT connected today's energy disaggregation paradigm, end-users will have more autonomy by receiving actionable insights by knowing the individual appliance behaviors and patterns resulting from the data-driven disaggregation mechanism. To 
explain the modeling of energy disaggregation, consider 
the following example with the aggregate data of energy appliances such as kettle, electric heater, laptop and projector. Suppose there is a sequence of readings from a house-level meter denoted as $X = (x_1, x_2, ..., x_t)$, where $t$ is the length of the sequence. The problem of energy disaggregation is to disaggregate $X$ into the energy consumption sequence of individual appliances:
\be 
Y^{i}=(y_{1}^{i},y_{2}^{i},...,y_{t}^{i}), y_{t}^{i}\in \uR^+,\ i = 1,\ldots, I,
\ee 
where $\uR^+ =[0,\infty)$, $I$ is the number of known appliances, and $t \in {1,2,..., T}$ is the index of samples.
At any time, $t$, $x_{t}$ is assumed to be the summation of the readings from all the known appliances and unknown appliances with background noise, denoted by $u_t$, i.e.,
\be 
X_{t} = \sum_{i=1}^{I}{y_{t}}^{i} + u_{t}.
\ee

Then, the energy disaggregation is to find the following inverse functions so that the energy consumption of individual appliances from the total power consumption can be obtained:
\be 
f^{i}: X \rightarrow Y^{i}, \ i = 1,\ldots, I.
\ee 
Since the energy disaggregation can identify and classify the energy consumption of individual appliances, it thus greatly facilitates energy control. Disaggregated data provides valuable information to facilitate power system planning, load forecasting, new types of billing procedures, and better service to customers (e.g., providing energy consumption details and pinpointing the origins of certain customer complaints). For policymakers, knowing the amount of energy each category of appliances consumes is critical to developing and evaluating energy  use policies. There are many applications of energy disaggregation, such as   accurate energy billing, occupancy monitoring, appliance classification, faulty appliance detection, building efficiency, and  demand-side management~\cite{NAJAFI2018377}.

There are two types of methods for energy disaggregation: Non-intrusive Load Monitoring and Intrusive Load Monitoring. `Non-intrusive' load monitoring consists of measuring energy consumption using a smart meter, which is usually placed at the meter panel and works based on the aggregated data where the disaggregation algorithms attempt to approximate the individual appliance characteristics. On the contrary, `Intrusive' load monitoring utilizes a low-end metering device for the measurement of the electricity consumption of one or more appliances residing within the end-user dwelling as indicated by the name~\cite{verma2021comprehensive}. While traditionally research has  focused on the non-intrusive energy disaggregation, recent advancement on the ML/artificial intelligence (AI), IoT/edge computing with the proliferation of accurate and affordable smart plugs have highlighted the need for research and development of the intrusive load monitoring.  

\subsection{Algorithms for Energy Disaggregation}
One of the most well-known energy disaggregation algorithm is the Hidden Markov Model (HMMs) series  \cite{eddy1996hidden}, such as Factorial Hidden Markov Models (FHMMs) and its variants like Additive Factorial Hidden Markov Models (AFHMMs). These HMM-based approaches work in a supervised or unsupervised setting, and their learning processes often rely on the expectation-maximization algorithm, thereby achieving the local-optima solutions. Other algorithms are Sparse Coding \cite{elhamifar2015energy}, Graph signal processing (GSP) \cite{stankovic2014graph}, Integer Programming and so on. As mentioned earlier, energy disaggreagtion is a special case of blind source separation, sparse coding has been proven effective for such problems where an additional constraint of sparse activations is introduced~\cite{Jiang21}. 

GSP utilizes the regularization of graph signals with the assumption that the signal is piecewise-smooth. Under this assumption and consideration, the total graph variation is generally not significant and can be used for a variety of applications including energy disaggregation. GSP has been proven effective over other algorithms because of its trainning efficiency, reliability in dealing with noisy data, ability to deal with different sampling rate, etc~\cite{HOLWEGER2019100244}. GSP-based approaches need more investigation related to the robustness of the algorithm, capability of dealing with incomplete training set, and real-time performance improvement. 

In the literature, ML based solutions have shown promising outcomes, especially, including all kinds of supervised classification algorithms, such as Random Forest and KNN classifier. Although the traditional algorithms have shown  promising outcomes, still there is scope for further accuracy improvement of the disaggregation algorithm. 

To this end, deep learning based methods, such as LSTM and autoencoder models with deep layers, have been shown effective for energy disaggregation algorithms. Kelly and Knottenbelt \cite{kelly2015neural} show that deep learning based energy disaggregation methods outperform the traditional ML-based methods. Existing deep learning algorithms for energy disaggregation tasks include Convolutional Neural Network (CNN), Recurrent Neural Network (RNN), Autoencoders. As an example, Chen et al. \cite{chen2018convolutional} used convolutional sequence to sequence model whereas Zhang et al. \cite{zhang2018sequence} showed the effectiveness of a sequence-to-point paradigm for energy disaggregation. Note that, sequence-to-point based CNN method outperformed the sequence-to-sequence learning approach.

Another type of neural network is `WaveNet', which has no recurrent connections and therefore, it is much faster than the traditional RNNs~\cite{oord2016wavenet}. WaveNet has been used for energy disaggregation in \cite{8682543}, showing better performance compared with other HMM-based algorithms and is  promising for data from multiple sources, e.g., combined consumption, and weather data. 

\subsection{Energy Usage Pattern Analysis}
The purpose of the energy usage pattern analysis is to give occupants a clear insight to understand their daily behavior in consuming electricity by identifying the usage patterns of various appliances, which can enhance their awareness of power-saving and lead them to a sustainable and healthier lifestyle \cite{ahmadi2016framework}.

Event detection is an effective way for analyzing the energy-consumption pattern for household appliances, as people’s indoor activities and the use of electrical appliances are closely related\cite{jazizadeh2018embed}. Another and more straightforward way is to detect the switch status of these devices, namely, on and off. For most appliances, there is a fixed power value for start and operation, i.e., on-power threshold. An appliance is recognized as starting working only when its output power reaches the threshold. Since different appliances have different on-power threshold, identifying and classification for various appliances can be achieved based on this point. For example, the on-power threshold of a kettle is about 2000 watts, while a washing machine only needs 20 watts. Such problems can be implemented using regression-based learning methods and classification-based learning methods \cite{Jiang21}. 

\section{Public Datasets of Power Consumption}\label{sec:4}

Although an IoT network is able to collect power consumption datasets, it is always beneficial to exploit a large volume of public datasets in cloud for training models. Thus, in this section, we summarize the existing public datasets.

\begin{table*}[htbp]
\caption{Table 1: Comparison of several disaggregation datasets \cite{jazizadeh2018embed} }
\centering
\begin{tabular}{lllllllll}
\hline
Dataset  & Location & \begin{tabular}[c]{@{}l@{}}No.of\\ units\end{tabular} & Duration    & \begin{tabular}[c]{@{}l@{}}Aggregate\\ Rate\end{tabular} & \begin{tabular}[c]{@{}l@{}}Appliance\\ Rate\end{tabular} & Labeled & Year & Avalibility \\ \hline
REDD \cite{kolter2011redd}     & USA & 6 & 3-19 days & 1Hz\&15kHz & 1/3 Hz & No & 2011 & \url{http://redd.csail.mit.edu}\\ \hline
BLUED \cite{anderson2012blued}    & USA & 1 & 8 days    & 12kHz      & -      & Yes& 2012 & \url{http://portoalegre.andrew.cmu.edu:88/BLUED}\\ \hline
Smart* \cite{barker2012smart}    & USA & 3 & 3 months  & 1Hz        & 1Hz    & No & 2012 &\url{http://traces.cs.umass.edu/index.php/Smart/Smart}\\ \hline
AMPds \cite{makonin2013ampds}    & Canada& 1 & 1 year  & 1/60 Hz    & 1/60 Hz& No & 2013 & \url{http://ampds.org/}\\ \hline
iAWE \cite{batra2013s}     & India & 1 & 73 Days & 1Hz        & 1Hz    & No & 2013 & \url{http://iawe.github.io/}   \\ \hline
SustData \cite{pereira2014sustdata} & Portugal  & 50 & 4-17 months  & 50 Hz    & 2s/3s  & No & 2014 &\url{https://osf.io/2ac8q/}\\ \hline
UK-DALE \cite{kelly2015uk}       & UK   & 4 & 3-17 months & 16 kHz  & 1/6 Hz   &Yes &2015 &\url{https://data.ukedc.rl.ac.uk}\\ \hline
SmartSim \cite{chen2016smartsim} & -  & 1 & 7 Days  & -          & 1 Hz   & -  & 2016 &\url{https://github.com/klemenjak/smartsim}\\ \hline
REFIT \cite{murray2017electrical} & UK   & 20 & 2 years & -           & 8-10s       &Yes    & 2017 & \url{https://www.refitsmarthomes.org/datasets/}\\ \hline
EMBED \cite{jazizadeh2018embed}    & USA   & 3 & 14-27 Days& 12 kHz   & 1 Hz   & Yes& 2018 &\url{http://embed-dataset.org/}\\ \hline
SynD \cite{klemenjak2020synthetic}     & -    & 1 & 180 days& -       & 5 Hz   & - & 2020 &\url{https://github.com/klemenjak/synd/}\\ \hline
\end{tabular}
\end{table*}
\subsection{Data Format for Energy Disaggregation}
Generally, there are two types of sample data. (1) One is high-frequency sampling data, which is usually the collected current and voltage current waveforms. These raw data can be converted to the power metrics using Short-time Fourier Transform  (STFT) \cite{jazizadeh2018embed}, i.e., real and reactive power with a time series. See the equations below:
\begin{align*}
P_{k}(t) & = \left | I_{k}(t) \right | \cdot \sin(\theta(t)) \cdot \left | V_{1}(t) \right | \\ 
Q_{k}(t) & = \left | I_{k}(t) \right | \cdot \cos(\theta(t)) \cdot \left | V_{1}(t) \right |
\end{align*}
in which $P_{k}$ is the real power, $Q_{k}$ is the reactive power, $I_{k}$ is the $k^{th}$ harmonic component of the transformed current waveform, and $V_{1}$ is the first harmonic component of the transformed current waveform. Note that the current and voltage sampling rates are usually 12kHz, while the processed power metrics are 60Hz. (2) Another is low-frequency sampling data, collected from individual smart plugs, smart meters,  ambient sensors (e.g., light intensity sensors \cite{ahmadi2016framework}), or WiFi modules. The plug load sampling rate is usually 1$\sim$2 Hz. 

Datasets with event labels would be very beneficial for energy disaggregation, as they can serve as a benchmark for model evaluation. However, collecting labelled datasets is not easy because the labelling process is often very costly, which requires a lot of human resources, resources, and time, including the need for  pre-installing sensors or electricity meters to collect raw data, recording events to mark the state changes of the appliance on the electricity consumption time series, taking a long time to record enough events for analysis. In \cite{ahmadi2016framework}, researchers have to manually label the detected event according to occupants' written diaries, which record daily activities and related electrical devices with timestamps. It can be said that  collecting labelled data used to be one of the biggest challenges for energy management. Therefore, some researchers are dedicated to collecting, labelling and publishing datasets for energy disaggregation - today, we have some public datasets available for direct use~\cite{anderson2012blued}.

\subsection{Power Consumption Dataset in the Public Domain}
There are some existing studies for energy disaggregation, which adopt supervised Machine Learning methods \cite{cunado2019supervised} to pursue high accuracy. That is, they rely on the labelled dataset with device signatures for training the model, and then the trained model can be used for energy disaggragation of the unlabelled data. Some studies are focusing on collecting dataset for energy monitoring \cite{kolter2011redd}\cite{anderson2012blued}\cite{barker2012smart}\cite{makonin2013ampds}\cite{batra2013s}\cite{pereira2014sustdata}\cite{kelly2015uk}\cite{chen2016smartsim}\cite{jazizadeh2018embed}\cite{klemenjak2020synthetic}. These datasets are compared in Table 1, and some public datasets available for researchers to develop new algorithms or as a test benchmark. 

The release of these datasets, especially some large-scale labelled datasets, accelerate the development of many ML algorithms for energy disaggregation and power consumption analysis. For example, Neural Networks and Deep Learning have been proven to outperform the previous algorithms (e.g., HMM) \cite{Jiang21}. These datasets with detailed labels can be used as training datasets, testing datasets and benchmarks to train ML models, evaluate model's performance, and compare energy-disaggregation algorithms. 

\section{Use-cases and Open Issues}\label{sec:5}

\subsection{Use-cases}

Energy disaggregation as a result of either intrusive or non-intrusive load monitoring has significant use-cases. It includes both energy efficiency and demand-side management while help the consumers to ensure cost-effective operation of their utilities. Some prospective use cases are accurate energy billing, occupancy monitoring, appliance classification, faulty appliance detection, building efficiency, demand-side management, etc, as discussed below~\cite{NAJAFI2018377}.

\subsubsection{Energy saving recommendations} Energy disaggegation will break down  energy consumption based on  individual appliances. Hence, looking at the consumption patterns or behaviors, a recommender system can  guide the non-expert towards optimal operation and cost savings.    
\subsubsection{Occupancy Monitoring} Disaggregated energy patterns will help to monitor the occupancy probability remotely. This use-case related to occupancy monitoring by activity recognition can be of great use for critical applications in age-care and medical facilities, children monitoring or remote occupancy identification in conference/meeting rooms, etc. 
\subsubsection{User psychology and cognitive behavior analysis}
    Energy usages patterns tell a lot about the habits and behavior of an individual. Authors in~\cite{Alcal2015DetectingAI} employs the energy disaggregation algorithms to detect the abnormal user activities within aging population. 
\subsubsection{Appliance classification} One of the most common usecases of energy disaggregation is  appliance classification. Future research aims to extend the capabilities to accurately identify different models of each appliance categories, their possible energy ratings, and other properties. 
\subsubsection{Faulty device diagnosis} Disaggregated energy profiling of appliances can help early identification of faulty devices. If one or more functionalities of any device is not working properly (e.g., device is not going in hibernation or sleeping mode), it will be captured by its individual consumption profile. Remedies taken from such action can help to diagnose the technical problem and long time cost savings.
\subsubsection{Context-aware smart systems} Energy profiling can help in providing additional context information for smart systems that can react and respond to changing situations in the home or office.

\subsection{Open Issues}

\subsubsection{Standardization of dataset}
If a new appliance or device is introduced, it may take some time to collect its power consumption data and train the models. For this, a local home server may use a dataset from the manufacturer  of the appliance if it is available. However, this requires standardization of datasets   so that the local home server can easily utilize the available dataset rather than collect and re-train locally. 

\subsubsection{Edge computing}
Although each local home server can train its own models, its computing power as well as datasets might be limited. To overcome those limitations, edge servers can be used. Each edge server can support a number of homes and offices simultaneously. In this case, each home/office needs to upload their datasets to an edge server. Unfortunately, this may cause privacy issues as home owners may not want to reveal their home devices and appliances together with their usage information. To avoid this problem, we can use secure communications \cite{Lee19} \cite{Lee20}. In addition, if the training models are shared by multiple 
homes/offices, 
federated learning  \cite{Li20} can be used to train the models, as each home/office can keep their datasets and only exchanges the parameter vectors of the models that share with other homes and offices.

\section{Conclusions}\label{sec:6}
This paper systematically reviews studies related to smart energy management based on the IoT network, focusing on state-of-the-art algorithms for energy/power disaggregation and public datasets of power consumption. We discussed potential use cases for smart energy management with future prospects and practical implications. Some challenges and open issues have also been  presented.  

\bibliographystyle{ieeetr}
\bibliography{ref}
\end{document}